\documentclass[12pt,epsfig]{article}
\usepackage{amssymb}

\def\beq{\begin{equation}}
\def\eeq#1{\label{#1}\end{equation}}
\def\eeqn{\end{equation}}

\def\beqa{\begin{eqnarray}}
\def\eeqa#1{\label{#1}\end{eqnarray}}
\def\eeqan{\end{eqnarray}}

\def\NPB{ Nucl. Phys. {\bf B}}
\def\PLB{Phys. Lett.  {\bf B}}
\def\PRL{Phys. Rev. Lett.}
\def\PRD{Phys. Rev. {\bf D}}

\let\bar=\overbar

\def\Dslash{\not{\hbox{\kern-4pt $D$}}}
\def\dslash{\not{\hbox{\kern-2pt $\del$}}}

\def\msb{{\bar{\ssstyle M \kern -1pt S}}}

\def\Title#1{\begin{center} {\Large #1 } \end{center}}

\begin{document}

\Title{TEN YEARS OF PRECISION ELECTROWEAK PHYSICS}

\bigskip\bigskip


\begin{raggedright}  

{\it A. Sirlin\index{Author, A.B.}\\
Department of Physics \\
New York University,
New York, 10003}
\bigskip\bigskip
\end{raggedright}

 
\section{Introduction and Brief Historical Perspective} 
\noindent
This summer we mark the tenth anniversary of the beginning of a very important
period in the study of the Standard Model (SM) \cite{WSG} and some of its
extensions.
Namely, in August of 1989, LEP started operations at CERN. Approximately at
the same time, SLC and the Mark II detector were switched on at SLAC, and
FNAL began the precision studies of the W mass. There have also been  
important contributions from other great laboratories. \\
A very attractive feature of this subject and period has been the
detailed interplay between theory and experiment. On the experimental
side, the accuracy often reaches $0.1 \%$ and sometimes it is much
better, as in the measurement of the Z mass. On the theoretical side,
the study of electroweak corrections to allowed processes, i.e.
processes not forbidden in lowest order, has been the
basis for the detailed comparisons currently achieved.\\
Since the path to renormalization has been often reviewed, I will
focus on certain aspects and applications of the theory that are
closely connected with experiment. This subject is vast and my time
 limited, so that 
the topics and references are not intended to be
exhaustive. In particular, apologies are due beforehand for the
omission of many important contributions.\\
From the beginning, in order to regularize ultraviolet divergences, most
calculations in electroweak physics have been carried out in the
dimensional regularization scheme \cite{tHVBG}. The application of this
approach to regularize infrared divergences and mass singularities was
proposed and analyzed somewhat later \cite{MGMSM}. In the seventies,
radiative
corrections to $\beta$ and muon decays played an important role in the
analysis of the universality of the weak interactions and its
implications for the phenomenological viability of the SM \cite{AS74}. The
evaluation of the one-loop corrections to $g_{\mu} - 2$ dates from
that period, and there were also a number of important qualitative
results, such as the absence of parity and strangeness violating
corrections of $O(\alpha)$ to strong interactions \cite{W73}, the
cancellation of ultraviolet divergences in natural relations \cite{BGS}, the
discovery that heavy particles do not generally decouple in
electroweak corrections and that a heavy top quark gives contributions
of $O(G_{\mu} m_t^2)$ to the $\rho$ parameter \cite{V77}, and the suppression
of flavor changing neutral currents (FCNC) in $O(G_f\alpha)$ \cite{LG}.
However, aside from the
problem of universality, it was too difficult in the seventies to
carry out sufficiently complete calculations of allowed processes that
could be compared with experiment at the loop level. One of the main
reasons was that, in order to establish connection with experiment, it
is generally necessary to evaluate corrections to several
processes. This important objective was hampered, partly by the large
number of diagrams involved, partly by the complicated nature of the
renormalization procedures frequently employed. Since 1980, simple
methods to implement the renormalization of the theory have been
developed.  These simple renormalization frameworks
have greatly facilitated the systematic evaluation of the electroweak
corrections to important allowed processes, such as muon decay, deep
inelastic neutrino-nucleon scattering, neutrino-lepton scattering,
atomic parity violation, and $e^{+} e^{-}$ annihilation at the Z peak
and in the LEP-II domain. In this way, the connection between theory
and experiment was finally achieved for a large number of
observables. In the early eighties, the main objective was the
prediction of $m_W$ and $m_Z$, using $\alpha$, $G_{\mu}$, and the
electroweak-mixing parameter
$\sin^2\theta_{W}$, as inputs. This required the corrections to muon
decay \cite{AS80}, and deep inelastic neutrino-nucleon scattering via the
neutral \cite{MS80} and charged currents \cite{MS81}. The measurement of
 $\sin^2\theta_W$
improved with time and by 1987 the calculated vector boson masses were
 $m_W =
80.2\pm1.1\, GeV$ and $m_Z = 91.6\pm0.9 \, GeV$ \cite{A87}, with central
values within 0.2 GeV and 0.4 GeV from the current ones, respectively.\\
A major breakthrough took place
with the onset of LEP. By the end of August of 1989, $m_Z$ had been
measured to within 160 MeV. It became immediately possible to obtain a
rather precise value of the $\overline{MS}$ parameter
$\sin^2\hat{\theta}_W(m_Z)$, which in turn helped to verify the
consistency with supersymmetric unification \cite{AS89}.  The precision of the
$m_Z$ measurement prompted a change in strategy: $\alpha$, $G_{\mu}$,
and $m_Z$ were adopted as the basic input parameters, a major effort
was placed on the study of the observables at the Z resonance, namely
the line shape and the various widths and asymmetries measured at LEP
and SLC, and there was a great improvement in the comparison between
theory and experiment. The main objectives of these studies have been:
 i) to test the SM at the level of its quantum corrections ii) to predict
$m_t$ and constrain the great missing piece, $M_H$ iii) to search for
deviations that may signal the presence of new physics beyond the
SM.\\
A good example of the successful interplay between theory and
experiment was the $m_t$ prediction and its subsequent
measurement. Before 1994-95, the top quark was not observed, but $m_t$
could be inferred indirectly because of its effect on electroweak
corrections. In Nov. 1994, a global analysis by the Electroweak
Working Group (EWWG) led to the indirect determination $m_t =
178\pm11^{+18}_{-19}$ GeV, where the central value corresponds to
$M_H$ = 300 GeV, the first error is experimental, and the second
uncertainty reflects the shift in the central value to $M_H$ = 65 GeV
and $M_H$ = 1 TeV. The present experimental value is $m_t =
174.3\pm5.1$ GeV; thus, we see that the indirect determination
was in the ball-park. Recent important results include precise measurements
 of $m_W$ at CDF, D0, LEP2, and NuTeV.

\section{Input Parameters}
\noindent
Three accurate input parameters play a major role in Electroweak
Physics.\\
1) $\alpha = 1/137.03599959(38)(13)$, most precisely derived from
$g_{e} - 2$, with an uncertainty $\delta\alpha$ =  0.0037ppm 
\cite{CZARNECKIMAR}. \\
2)~$G_{\mu} = (1.16637\pm0.00001)\times10^{-5} GeV^{-2}$. It is
derived from the muon lifetime with an uncertainty of 9ppm, using
the radiative corrections of the V-A Fermi theory. Recently, the
two-loop corrections to the muon lifetime have been completed in the 
approximation of neglecting terms of $O(\alpha^2 (m_e/m_{\mu})^2)$
\cite{RSS}. Using $\alpha$ as expansion parameter, one has:
\begin{displaymath}
\delta = 1+  \frac{\alpha}{2 \pi} \left( \frac{25}{4}  -
 \pi^2 \right) \left[ 1 + \frac{2 \alpha}{3 \pi}
 \ln \left( \frac{m_{\mu}}{m_e} \right) \right] +
 6.701 \left( \frac{\alpha}{\pi} \right) ^2 + \ldots
\end{displaymath}
\noindent
The contributions of $O(\alpha)$  and $O(\alpha^2 \ln(m_{\mu}/m_e))$
 have been known for a long time \cite{KSBR}, while the last term is the new
result. The two $O(\alpha^2)$ terms nearly cancel and, including very
small $O(\alpha m_e^2 / m_{\mu} ^2 ) $ contributions, one finds:\\
\begin{displaymath}
\delta = 1 - 4.1995 \times 10^{-3} + 1.5 \times 10^{-6} + \ldots
\end{displaymath}
\noindent
3) $m_Z = 91.1872\pm0.0021 GeV$ with an uncertainty of
23ppm. It turns out that there are subtleties in the theoretical
definition of $m_Z$ and the width $\Gamma_Z$, associated with issues of
gauge invariance and, in the case of photonic and gluonic corrections
to $W^{\pm}$ and quark propagators,
with questions involving the convergence of the perturbative expansion
in the resonance region \cite{S9198S} . Thus, the study of this region
 sheds light on the concepts of mass and width of unstable particles!

\section{ \boldmath{$m_W$}, \boldmath{$\sin^2 \theta_W$}, 
On-Shell and \boldmath{$\overline{MS}$} Renormalization
 Schemes}
\noindent
Other fundamental parameters are $m_W$ , the physical (pole) mass of
the W boson and the electroweak mixing parameter. The latter comes in
several incarnations, all of them interesting.\\ $\sin^2 \hat{\theta}
_W (m_Z)$ = $\hat{s} ^2$ is the renormalized parameter in the
$\overline{MS}$ renormalization scheme, evaluated at the $m_Z$
scale. A frequently employed version of this method, applied in the
early 80's \cite{MSSMS}, and further developed  since
1989 \cite{AS89,DFS91,DEGSIR,FKS} employs $\overline{MS}$ couplings and
 physical
(pole) masses
in the electroweak sector. $\hat{s} ^2$ is very convenient to discuss
physics at the Z peak, and is crucial for GUTs studies.\\ $\sin^2
\theta_W = 1 - m_W^2 / m_Z^2 $ = $s^2$ is the renormalized parameter in the
on-shell method of renormalization, developed since 1980 \cite{AS80,MS80,AS84,
H90}. It employs physical parameters such as $\alpha$, $m_W$, $m_Z$, $G_{\mu}$
in the electroweak sector.\\
Using $\alpha$, $G_{\mu}$, and $m_Z$ as inputs, one can evaluate $m_W$ and
$\hat{s}^2$, as functions of $m_t$ and $M_H$. One has the relations:\\   
\begin{displaymath}
s^2 c^2 = \frac{A ^2}{m_Z^2 (1 - \Delta r)} \quad ; 
\quad   \hat{s}^2 \hat{c}^2 =
\frac{A^2}{m_Z^2 (1 - \Delta \hat{r})} \quad ; 
\quad      \hat{s}^2 = \frac{A^2}{m_W^2 (1 - \Delta \hat{r}_W )}\quad ,\\
\end{displaymath}
where $A^2 = \pi \alpha / \sqrt{2} G_{\mu}$. The corrections $\Delta
r$ \cite{AS80}, $\Delta \hat{r}$ \cite{AS89,DFS91}, and
$\Delta \hat{r}_W$ \cite{DFS91} 
play an
important role in the analysis of electroweak physics, because they
link $\alpha$, $G_{\mu}$, and $m_Z$ to the precisely determined
parameters $m_W$ and $\hat{s}^2$. In particular, as it is clear from
the first equation, $\Delta r$ is a physical observable. Therefore, it
can be evaluated in any renormalization scheme.\\ The $\overline{MS}$
and on-shell definitions of the electroweak mixing angle are related
by\\
\begin{displaymath}
\hat{s} ^2 = s^2 \left( 1 + \frac{c^2}{s^2} \Delta \hat{\rho} \right) \quad ;
\quad
\Delta \hat{\rho} = Re \left( \frac{ A_{WW} (m_W^2) }{ m_W ^2 } -
\frac{ A_{ZZ} (m_Z^2) }{ m_Z^2 \hat{\rho} } \right) ,\\
\end{displaymath}
where $A_{WW}(m_W^2)$ and $A_{ZZ}(m_Z^2)$ are the W-W and Z-Z self-energies
evaluated on their mass-shells, and renormalized in the $\overline{MS}$ scheme
at the $m_Z$ scale,
and $\hat{\rho} =
c^2 /  \hat{c}^2 = \left( 1 - \Delta \hat{ \rho} \right) ^{-1}$.\\
The neutral current amplitude is of the form\\
\begin{displaymath}
<f \bar{f}| J^{z}_{\mu} | 0> = V_f (q^2) \bar{u}_f \gamma_{\mu} 
\left[ \frac{I_3 ( 1 - \gamma_5 )}{2} - \hat{k}_f (q^2) \hat{s}^2 Q_f \right]
v_f,\\
\end{displaymath}
where $\hat{k}_f (q^2)$, its on-shell counterpart $k_f (q^2) =
\hat{k}_f (q^2) \hat{s}^2 /s^2$, and $V_f (q^2)$ are electroweak form
factors, $I_3$ is the third component of weak isospin, and $Q_f$ is
the charge of fermion f. A dominant contribution to $\Delta
\hat{\rho}$ can be identified with the top quark contribution to the
$\rho$ parameter:
\begin{displaymath}
\Delta \rho_t = \frac{3 G_{\mu} m_t^2}{ 8 \pi^2 
\sqrt{2}}(1 - 0.12) = 8.4 \times 10^{-3},\\
\end {displaymath}
where the last factor represents the QCD correction. Similarly, if the
neutral current amplitude is parametrized in terms of $G_{\mu} m_Z^2$,
it contains a contribution proportional to $(1 - \Delta \rho_t)^{-1}
$. However, $\Delta\hat{\rho}$ includes gauge-invariant bosonic 
contributions and the self-energies in $\Delta \rho_t$ are evaluated at
$q^2 = 0$, so that there are significant conceptual and numerical 
differences between the two corrections.\\
Another important definition is $\sin^2\theta^{lept}_{eff} =
s^2_{eff}$, used by the EWWG to parametrize the data at the Z
peak. It is related to the other definitions by  $s^2_{eff} =
Re\hat{k}_l(m_Z^2)\, \hat{s}^2 = Rek_l(m_Z^2)\, s^2$ \cite{GS}, where
 $\hat{k}_l$
and $k_l$ are the electroweak form factors in the f=lepton case. By a
fortuitous cancellation of electroweak corrections, $\hat{k}_l
(m_Z^2)$ is very close to 1 and, for current $m_t$ values, $s^2_{eff}
- \hat{s}^2 \thickapprox  1 \times 10^{-4}$. Writing $\hat{k}_l = 1 + \Delta
\hat{k}_l$, and taking into account the smallness of $\Delta
\hat{k}_l$, the combination with the expression for $\hat{s}^2
\hat{c}^2$ leads to
\begin{displaymath}
s^2_{eff} c^2_{eff} = \frac{A^2}{m_Z^2(1 - \Delta r_{eff})} \quad ; \quad
\Delta r_{eff} = \Delta \hat{r} + \left(1 - \frac{\hat{s}^2}{\hat{c}^2}\right)
Re \Delta \hat{k}_l.\\
\end{displaymath}
As $\Delta r$, $\Delta r_{eff}$ is scale independent, since it is defined
in terms of observable quantities.\\
Other interesting renormalization schemes include the formulation presented
in Ref. \cite {Nov}. However, the on-shell and $\overline{MS}$ renormalization
schemes remain the most frequently employed, within the SM and beyond. For
example, the ZFITTER  and  BHM
programs are
 based
on the on-shell method, while the GAPP \cite{GAPP} and TOPAZ0 
codes employ the $\overline{MS}$ formulation.

\section{The Running of \boldmath{$\alpha$}. Asymptotic Behavior of Basic Corrections}
\noindent
An important contribution to the basic electroweak corrections is
associated with the running of the QED coupling at the $m_Z$ scale:
$\alpha (m_Z)/ \alpha = 1/ (1 - \Delta \alpha )$. The contribution of
the light quarks (u through b) is evaluated using dispersion
relations  and $\sigma_{exp}(e^{+}+e^{-}
\to hadrons)$ at low $\sqrt{s}$ and perturbative QCD (PQCD) at large 
$\sqrt{s}$.
A frequently employed value is $\Delta \alpha^{(5)}_{h} = 0.02804
\pm0.00065$ \cite{EJ}. Recently, several ``theory driven'' calculations
claim to sharply reduce the error by extending the application of PQCD
to much lower $\sqrt{s}$ values ($\sqrt{s} \thickapprox 1.7 GeV$). An
example is $\Delta \alpha^{(5)}_{h}= 0.02770 \pm0.00016$
\cite{DH}. There is a new calculation that applies PQCD to the Adler function
$ D(Q^2) = Q^2 \int_{4m_{\pi}}^{\infty} ds' R(s')/(s' + Q^2)^2 $, down
to $\sqrt{Q^2} = 2.5 GeV$, where $Q^2 = -s$ is space-like \cite{J}. The
authors find $\Delta\alpha^{(5)}_{h}(-m_Z^2)$, evaluate the difference
with $\Delta\alpha^{(5)}_{h}(m_Z^2)$ using PQCD, and obtain
$\Delta\alpha^{(5)}_{h}(m_Z^2) = 0.027782\pm0.000254$. The
space-like approach circumvents possible problems associated with
time-like thresholds.\\
\noindent
The basic corrections have been studied in great detail by several
groups. It is instructive to display their asymptotic behaviors for
large $m_t$ and $M_H$:
\begin{displaymath}
\Delta r \thicksim -\frac{3\alpha}{16 \pi s^4} \frac{m_t^2}{m_Z^2} +
\frac{11 \alpha}{24 \pi s^2} \ln\left( \frac{M_H}{m_Z}\right) + \ldots \\
\end{displaymath}
\begin{displaymath}
\Delta r_{eff}  \thickapprox  \Delta\hat{r}   \thicksim  -  
\frac{3 \alpha}{16 \pi  \hat{s}^2 \hat{c}^2} \frac{m_t ^2}{m_Z^2} +
\frac{\alpha}{2 \pi \hat{s}^2 \hat{c}^2}  \left(\frac{5}{6}  - 
\frac{3}{4} \hat{c}^2 \right)  \ln\left(\frac{M_H}{m_Z}\right)  +  \ldots\\
\end{displaymath}
These formulae exhibit some of the main qualitative features of the
corrections: a quadratic dependence on $m_t$, enhanced by a relative
factor $c^2/s^2$ in $\Delta r$, a logarithmic dependence on $M_H$, and
opposite signs. The latter leads to a well-known correlation between
the $m_t$ and $M_H$ values derived from the electroweak data.
Variations of $\Delta r$ and $\Delta \hat{r}$ induce
shifts $\delta m_W / m_W \thickapprox - 0.22 \delta (\Delta r)$ and
$\delta s_{eff}^2 / s_{eff}^2 \thickapprox 1.53 \delta (\Delta
\hat{r})$.

\section{Evidence for Electroweak Corrections}
\noindent
We discuss two classes of interesting loop contributions.\\
A) Corrections Beyond the Running of $\alpha$.\\ A sensitive argument
is to measure $\Delta r$ \cite{AS94}.  Using the current world average $m_W =
80.394 \pm0.042 GeV$ \cite{EWWG} , one finds $(\Delta r)_{exp} = 0.03447
\pm0.00251$, while the contribution to $\Delta r$ from the
running of $\alpha$ is $\Delta\alpha = 0.05954\pm0.00065$.  Thus,
the electroweak correction not associated with the running of $\alpha$
is $(\Delta r)_{exp} - \Delta \alpha = -0.02507\pm0.00259$, which
differs from zero by $9.7 \sigma$!  If, instead, one employs the 
$m_W$ value from the global fit, which includes both direct and
indirect information, the evidence for corrections beyond the running
of $\alpha$ is close to the $14 \sigma$ level. \\
A similar result is obtained by comparing $s^2_{eff}$ and $\sin^2\theta_W$,
both of which are physical observables \cite{AS94}. Their numerical difference
arises from electroweak corrections not involving $\Delta\alpha$, and 
amounts to $0.00879\pm 0.00083$, a 10.6 $\sigma$ effect.\\
B) Electroweak Bosonic Corrections (EBC).\\ These include loops
involving the bosonic sector, W's, Z, H. They are sub-leading
numerically, relative to the fermionic contributions, but very
important conceptually. Strong evidence for these corrections can be
obtained from $\Delta r_{eff}$ \cite{GSEC}.  Using the current average
$s^2_{eff} = 0.23151 \pm0.00017$, one finds $(\Delta
r_{eff})_{exp} = 0.06052 \pm0.00048$. Subtracting the EBC
diagrams ( a gauge invariant and finite subset) from the theoretical
 evaluation,
one obtains  $(\Delta r_{eff})^{subtr}_{theor} = 0.05106\pm
0.00083$. The difference is $0.00946\pm0.00096$, a $9.9 \sigma$
effect!

\section{Theoretical Pursuit of the Higgs Boson}
\noindent
With $m_t$ measured, the question of whether and to what extent it is possible
to constrain
 $M_H$ , becomes of considerable interest.  One faces the
problem that asymptotically the electroweak corrections are $\thicksim
\ln(M_H/m_Z)$, so that precise calculations are necessary. It is therefore
important to consider the level of accuracy of the electroweak corrections.\\
Theorists distinguish two classes of
errors: 1) parametric 2) uncertainties due to the truncation of the
perturbative series (i.e. un-calculated higher order effects). The
first class includes the errors in $m_Z$, $\Delta\alpha^{(5)}_{h}$,
$ m_t$, $ s^{2}_{eff}$, $ m_W$, and $ \alpha_{s}(m_Z^2)$.  
In principle, they can be decreased by improved
experiments. The second class is more difficult to estimate.  What is
the current theoretical situation? Corrections of $O(\alpha)$,
 $O([\alpha \ln(m_Z/m_f)]^n)$, and $O( \alpha^2 \ln(m_Z/m_f))$, where
$m_f$ is a generic light-fermion mass, were analyzed
from around 1979 to 1984 \cite{AS84,M79}. Those of
 $O(\alpha^2 (M^{2}_{t}/ M^{2}_{W})^2)$ \cite{CJHBARFL}, as well as the
 QCD corrections of
 $O(\alpha\alpha_s)$ \cite{FKS,K90} and
 $O(\alpha \alpha^{2}_{s} (M^{2}_{t} / M^{2}_{W}))$ \cite{AVCH}, 
were studied from the late eighties to the middle nineties.\\
Of more recent vintage is the analysis of the corrections of
 $O(\alpha^2 m_t^2/m_W^2)$. Large $m_t$ expansions were employed to
evaluate the irreducible contributions of this order to $\Delta r$ and 
$\Delta\hat{r}_W$ 
in the framework of the $\overline{MS}$ renormalization scheme \cite{DGV}.
In order to estimate the theoretical error due to the truncation of the
perturbative series, it was very useful to carry out analogous calculations
in other schemes, such as the on-shell renormalization framework. It was
also important to incorporate these effects in the theoretical 
evaluation of $s^{2}_{eff}$. In order to achieve these goals,
the corrections of $O(\alpha^2 m_t^2/m_W^2)$ in the calculation of $m_W$, 
$s^{2}_{eff}$, and $\hat{s}^2$ were incorporated and compared, as functions
of $M_H$, in three
schemes : $\overline{MS}$,  and two versions , OSI and OSII,  of the
on-shell scheme, with two different implementations of the
QCD corrections \cite{DGS}. A large reduction was found in the scheme
dependence. The maximal differences among the three calculations, for given
 $M_H$, amounted to $\Delta s^{2}_{eff}\thickapprox 3 \times 10^{-5}$ and
$\Delta m_W \thickapprox 2 MeV$ while, without the incorporation of the
new corrections, the variations were $\thickapprox 2 \times 10^{-4}$
and $\thickapprox 11$ MeV, respectively. Including additional  
QCD uncertainties, the estimated errors became
 $\Delta s^{2}_{eff} \thickapprox 6 \times 10^{-5}$,
 $\Delta m_W \thickapprox 7 MeV$ \cite{DGPS}.\\
The $O(\alpha^2 m_t^2/m_W^2)$ corrections in the
calculation of the partial widths $\Gamma_f (f\neq b)$ was studied 
in Ref. \cite{DG}, and a partial check of the accuracy of the Heavy
Top Expansion was carried out in Ref. \cite{GSW}.\\       
The incorporation of the $O(\alpha^2 m_t^2/m_W^2)$ contributions had a
felicitous consequence: it led, for equal inputs, to a significant
reduction in the derived value of $M_H$ and its upper bounds. For example,
a fit to the data by the EWWG , without inclusion of these effects,
led in Aug. 1997 to $M_H < 420 GeV$ at
$95\%$ CL, while Ref. \cite{DGPS}, using the
same input values, reported
$M_H < 295 GeV$, a $30\%$ reduction! The
$O(\alpha^2 m_t^2/m_W^2)$ contributions of Refs. \cite{DGV,DGS} have
been incorporated for some time in the Erler-Langacker analysis and, more
recently, into the ZFITTER and TOPAZ0 codes.

\section{Simple Formulae for \boldmath{$\sin^2 \theta^{lept}_{eff}$} and
\boldmath{$m_W$}}
\noindent
It turns out that simple and accurate formulae for
$\sin^2 \theta^{lept}_{eff}$ and $m_W$,
as functions of $M_H$, $m_t$, $\Delta\alpha^{5}_{h}$, and $\alpha_s (m_Z)$,
are available in the $\overline{MS}$, OSI, and OSII schemes \cite{DGPS}.
For example, in the $\overline{MS}$ framework, one has:
$$
\sin^2 \theta^{lept}_{eff} = 0.231510 +
0.000523~ {\rm ln} \left(\frac{M_H}{100} \right) +
0.00986 \left(\frac{\Delta \alpha^{(5)}_{h}}{0.0280} - 1 \right)
$$
\begin{displaymath}
- 0.00278 \left ( \left(\frac{m_t}{175}\right)^2  -1 \right) +
 0.00045 \left(\frac{\alpha_s(m_Z)}{0.118} - 1 \right),\qquad {\rm (I)}
\end{displaymath}
$$
m_W = 80.3827 - 0.0579\, {\rm ln} \left(\frac{M_H}{100} \right) -
0.008~ {\rm ln}^2 \left(\frac{M_H}{100} \right) -
0.517 \left(\frac{\Delta \alpha^{(5)}_{h}}{0.0280} - 1 \right) 
$$
$$
+~ 0.543~\left ( \left(\frac{m_t}{175}\right)^2  -1 \right) -
0.085~\left(\frac{\alpha_s(m_Z)}{0.118} - 1 \right),\qquad {\rm (II)}
$$
where $m_t$, $M_H$, and $m_W$ are expressed in GeV units. These formulae
reproduce accurately the detailed numerical results obtained in
Ref. \cite{DGS} in the range $75 \leqq M_H \leqq 350 $ GeV, with
maximum absolute deviations of $(1.1 - 1.3)\times 10^{-5}$ in the case
of Eq.(I), and $(0.8-0.9)$ MeV in that of Eq.(II).\\
\noindent
As an illustration, using $s^2_{eff} = 0.23151 \pm0.00017$ \cite{EWWG},
 $m_t = 174.3 \pm5.1$
GeV \cite{EWWG}, $\Delta\alpha^{(5)}_{h} = 0.02804 \pm0.00065$ \cite{EJ},
and $\alpha_s(m_Z) = 0.119 \pm0.003$ \cite {EWWG} in Eq.(I), one finds
${\rm ln} (M_H/100)= -0.077 \pm0.629 $, which corresponds to a central
 value $M^{c}_{H} = 93\, GeV$ and a $95\%$ CL upper bound
 $M^{95}_{H} = 260\, GeV$.
Inserting this value of ${\rm ln} (M_H/100)$ in Eq.(II), one obtains
the accurate SM prediction $m_W = 80.381\pm0.028 GeV$,  to be compared with
the current world average $(m_W)_{exp} = 80.394\pm0.042 GeV$.\\
If, instead, one uses $m_W =80.394\pm0.042 GeV$ as input in Eq(II), 
one finds $M^{c}_{H} = 73\, GeV$ and $M^{95}_{H} = 294\, GeV$. Thus, we see
that the $m_W$ measurement already leads to constraints on  $M_H$ not
far from those derived from $s^2_{eff}$!

\section{Recent values of \boldmath{$m_W$} and \boldmath{$\sin^2\theta^{lept}_{eff}$}, and salient
results from global fits}
\noindent
Recent values of $m_W$ include the $p-\bar{p}$ average from CDF and
D0 : $m_W=80.448\pm0.062\,GeV$ and the LEP-II result $m_W=80.350\pm0.056\,
GeV$. The difference between these two measurements is $1.2\sigma$ and their 
average is $m_W = 80.394\pm0.042\,GeV$. The NuTeV/CCFR value, whose
extraction depends weakly on $m_t$ and $M_H$, is $m_W=80.25\pm0.11\,GeV$.\\
In order to discuss $s^2_{eff}$, it is convenient to recall  the 
parity-mixing amplitude $A_f = 2v_f a_f /(v_f^2 + a_f^2)$,
where $v_f$ and $a_f$ are the vector and axial-vector couplings of
fermion f with the Z boson at resonance, and the very useful formulae:
$A^{o,f}_{FB} = (3/4)\,A_e\,A_f$; $A_{LR} = A_e$;
$A^{FB}_{LR}(f)= (3/4) A_f$.  This summer, a new value 
has been reported for the leptonic amplitude $A_l$ at SLD:
$A_l(SLD) = 0.1512\pm0.0020$, which implies 
 $s^2_{eff}=0.23099\pm0.00026$. The combined LEP and SLD value of $A_l$
is $A_l(LEP+SLD)=0.1497\pm0.0016$, which corresponds to
 $s^2_{eff}=0.23119\pm0.00020$ (only leptonic amplitudes). There has
also been some change in the forward-backward asymmetry $A^{o,b}_{FB}$
in the $b-\bar{b}$ channel. According to a recent analysis \cite{EWWG},
the new world-average is $s^2_{eff}=0.23151\pm0.00017$ with 
$\chi^2/{\rm d.o.f.} = 11.9/6$. Statistically, this corresponds to
a Confidence Level of $6.4 \%$, which is rather low. In particular, the
difference between the two most precise values of $s^2_{eff}$, derived from
$A_l(SLD)$ and $A^{o,b}_{FB}$, is 2.9 $\sigma$. A variation of the  same 
magnitude exists  between the ``leptonic'' and  ``hadronic'' averages
of $s^2_{eff}$,
involving $A_l$ and $A_q$, respectively. This can be compared with
$\chi^2/{\rm d.o.f.} = 7.8/6$ at the time of the Vancouver Conference last
year, with a statistical confidence level of $26 \%$. Thus, the fit to
this crucial parameter is less harmonious at present (summer of 1999) than 
it was last year.\\
We now present some salient results from two recent global fits.\\
The Erler-Langacker fit, which is based on the $\overline{MS}$ scheme,
leads to
\cite{GAPP} $\hat{s}^2 = 0.23117\pm 0.00016$, $m_t = 172.9\pm 4.6 GeV$,
$\alpha_s (m_Z) = 0.1192\pm 0.0028$. They employ a value of 
$\Delta \alpha^{(5)}_{h}$ which is adjusted in the fit and correlated with
that of $\alpha_s (m_Z)$, and a definition of $\hat{s}^2$ 
in which the contribution of the top quark is decoupled. In such a case
 the difference with $s^2_{eff}$ amounts to $2.9 \times 10^{-4}$
\cite{GS}, so that $s^2_{eff} = 0.23146\pm 0.00016$, a fit value somewhat
lower than the one derived directly from the various asymmetries measured
at LEP and SLC. Their indirect determination of $M_H$ is
\begin{displaymath}
M_H = 98^{+57}_{-38}\, GeV \qquad ; \qquad M^{95}_{H} = 235\, GeV.
\end{displaymath}
The $95\%$ CL upper bound $M^{95}_{H}$ takes into account
the exclusion constraint from the direct searches of H which, at the time,
was $M_H > 95 GeV$. This fit has
 a $\chi^2/{\rm d.o.f.} = 42/37 $, corresponding to a CL$\thickapprox 25\%$.\\
A second fit \cite{Quast}, uses the ZFITTER 
program employed by the EWWG, a code based  on the on-shell framework.~The
 value of
$s^2_{eff}$ is the
one mentioned before, $s^2_{eff} = 0.23151\pm0.00017$. Using 
  $\Delta \alpha^{(5)}_{h} = 0.02804 \pm0.00065$, it leads to 
$m_t = 173.2^{+4.7}_{-4.4}\, GeV$,~$\alpha_s (m_Z) = 0.1184\pm 0.0026$, 
$M_H = 77^{+69}_{-39}\, GeV$, and $M^{95}_{H} = 215\, GeV$, which does not
take into account the exclusion constraint from the direct searches.
Instead, employing $\Delta \alpha^{(5)}_{h} = 0.02784 \pm0.00026$, 
one of the recent theory driven calculations, the value of $m_t$ 
changes by only 0.2 GeV, $\alpha_s(m_Z)$ is unchanged, and   
$M_H = 90^{+57}_{-37}\, GeV$.\\
In general, the $m_t$ values from the global fits are smaller by $1-2 GeV$
from the direct measurements and, through the $m_t- M_H$ correlation
mentioned before, this tends to lower somewhat the $M_H$ value. The new
theory driven calculations of  $\Delta \alpha^{(5)}_{h}$ give a larger
central value for $M_H$ and a smaller error than the conventional
calculations. As a consequence, the upper bound $M^{95}_{H}$ turns out
to be rather close in the two cases. \\
The largest deviations between observables and  the values in the
Erler-Langacker fit are in $A^{o,b}_{FB} : -2.3 \sigma$ (LEP)~;
in the atomic parity-violation in Cs, $Q_W(Cs): 2.3 \sigma$~; in the
leptonic amplitude measured at SLC, $A_l: 1.8 \sigma$ (SLC)~; and in
the hadronic cross-section at resonance, $\sigma_{had}: 1.7 \sigma$ (LEP). 
(It is important to note that the theoretical error in $Q_W(Cs)$ has been
greatly reduced recently \cite{WIE}).
Thus, there are only two observables with deviations larger than $2 \sigma$
and two additional ones with variations close to $2 \sigma$.\\
However, if one combines  $A^{o,b}_{FB}$, $A_l(LEP + SLC)$, and $A_b (SLC)$,
one obtains a value for the b-quark amplitude $A_b$ which deviates from
the SM model prediction by $-2.7 \sigma$. The question has been raised
of whether this is due to a statistical fluctuation, or possible new physics
coupled to the third generation \cite{MAR,CHAN}. In particular, Ref. 
\cite{CHAN} discusses the implications of the new physics scenario for
FCNC. On the other hand, the analysis shows that, if the effect is due
to new physics, a substantial, tree-level  change in the right-handed 
$Z-b\bar{b}$ coupling is required.

\section{Vacuum Stability and Perturbation Theory Constraints. Supersymmetry}
In the very hypothetical scenario in which the SM is valid up to
energy scales $\Lambda \thicksim 10^{19}\, GeV$, one has the theoretical
inequality $134\, GeV\lesssim M_H \lesssim 180\, GeV$, where the lower
and upper bounds arise from the requirements of vacuum stability \cite{QUI}
and the validity
of Perturbation Theory, respectively \cite{RIE}. These estimates are for
 $m_t = 175\, GeV$ and $\alpha_s(m_Z) = 0.118$. If, instead,
$\Lambda \thicksim 10\, TeV$, the inequality becomes
$85\, GeV \lesssim M_H \lesssim 480\, GeV$.\\
Over the last several years, supersymmetric scenarios  have emerged as
leading candidates  for physics beyond the SM.\\
It has been known for a long time that the precision data is compatible
with SUSY grand-unification (SUSY GUTs) at
$\thicksim 2 \times 10^{16}\,GeV$! For instance, assuming supersymmetric
 unification of couplings, and using $\hat{\alpha}(m_Z)$, $\hat{s}^2 (m_Z)$
as inputs, one derives $\alpha_s(m_Z) = 0.13\pm0.01$, which is consistent
with current experimental values \cite{LANGPOL}.\\
A major prediction of the Minimal Supersymmetric Standard Model (MSSM) is
$m_h \lesssim 135 GeV$, where $m_h$ stands for the mass of the lightest CP-even
Higgs scalar. Recent diagrammatic calculations of $m_h$ include terms of
$O(\alpha \alpha_s)$ and significantly reduce the $\tan \beta$ region that
can be probed  by the Higgs boson search at LEP-II \cite{HEINE}.\\ 
In the MSSM,
SUSY contributions decouple if the superpartners' masses are much larger 
than $m_Z$.  In that 
regime, the fits are of the same general quality as in the case of the SM.
If some of them are of $O(m_Z)$, the fits are worse, which leads to
constraints in SUSY parameter space. In particular, both non-oblique
and oblique corrections are important in that case \cite{EP}.

\section{Constraints on Additional Fermions and Bosons; S,T,U parameters} 
Erler and Langacker \cite{GAPP} introduce a parameter
$\rho_o = c^2/(\hat{c}^2 \hat{\rho}_{SM})$, where $\hat{\rho}_{SM}$ is the
SM value of $\hat{\rho}$. It is sensitive to contributions from
non-degenerate additional doublets, and non-standard H bosons transforming
according to representations other than singlets and doublets. Fitting the
data with this additional parameter, they find 
$\rho_o = 0.9998^{+0.0011}_{-0.0006}$, $95\,GeV < M_H < 211 \,GeV$, 
$m_t = 173.6\pm 4.9\, GeV$, $\alpha_s(m_Z) = 0.1194 \pm 0.0028$, where the
lower $M_H$ limit was the direct search bound at the time. This fit is in
excellent agreement with the SM prediction $\rho_o = 1$. At the $2 \sigma$
level one has $\rho_o = 0.9998^{+0.0034}_{-0.0012}$ and 
$M_H < 1002 \,GeV$. Thus, in the presence of possible additional
contributions to $\rho_o$, the constraint on $M_H$ becomes very weak.
This analysis implies $\sum_{i} (C_i /3) (\Delta m_i)^2 \leqslant (100 GeV)^2$
at $95\%$ CL, where the sum is over additional fermion and scalar doublets
and $C_i = 1 (3)$ for color singlets (triplets). The bound is sharper
for non-degenerate squark and slepton doublets, namely $(69\,GeV)^2$,
on account of the strong correlation between $\rho_o$ and the 
restricted SUSY value of $m_h$.\\
The S,T, and U parameters are very useful to analyze possible
new physics contributions that reside in the self-energies (also called
oblique corrections) and which involve generic masses $m_i >> m_Z$
\cite{PESKIN}. One has
\begin{displaymath}
\hat{\alpha}(m_Z)\,T \equiv \frac{A^{new}_{WW}(0)}{m_W^2} -
\frac{A^{new}_{ZZ}(0)}{m_Z^2}\quad; \quad 
\frac{\hat{\alpha}(m_Z)}{4\hat{s}^2\hat{c}^2}\,S \equiv 
\frac{A^{new}_{ZZ}(m_Z^2) - A^{new}_{ZZ}(0)}{m_Z^2}
\end{displaymath}
\begin{displaymath}
\frac{\hat{\alpha}(m_Z)}{4\hat{s}^2}\,(S+U) \equiv 
\frac{A^{new}_{WW}(m_W^2) - A^{new}_{WW}(0)}{m_W^2},
\end{displaymath}
where the superscript new indicates that only new physics contributions
are included. They are part of the new physics contributions to the
basic corrections \cite{MARROS}. In fact, we have:
\begin{displaymath}
\delta(\Delta\hat{r})\thickapprox \delta(\Delta r_{eff}) = 
\frac{\alpha}{4\hat{s}^2\hat{c}^2}\,S - \alpha \,T \quad ;\quad
\delta(\Delta r_W) =\frac{\alpha}{4\hat{s}^2}\,(S+U)
\end{displaymath}
\begin{displaymath}
\delta(\Delta r) = \frac{\alpha}{2\hat{s}^2}\,S -
\frac{\alpha}{4\hat{s}^2}\left( \frac{c^2}{s^2} - 1 \right)\,U -
\alpha \frac{c^2}{s^2}\,T.
\end{displaymath}
For contributions involving masses $m_i >> m_Z$, the S and
S+U parameters are approximately proportional to the wavefunction 
renormalizations of the Z and W bosons, respectively. Heavy non-degenerate
additional doublets contribute positively to $ \hat{\alpha}\,T =   \rho_o - 1$, and to a lesser extent to U. The S parameter, instead, is sensitive 
to heavy degenerate chiral fermions, with a contribution $S = 2/3\pi$ per
generation. As loops involving the Higgs boson  affect mostly the
self-energies,
the S, T, and U parameters cannot be fitted simultaneously with $M_H$. 
Therefore, a value of $M_H$ is usually assumed. A recent fit \cite{GAPP} gives
$S= -0.07\pm0.11(-0.09), T= -0.10\pm0.14 (+0.09), U= 0.11\pm0.15(0.01)$.
The central values assume $M_H=100\,GeV$, while the values in parentheses 
indicate the change corresponding to $M_H=300 \,GeV$. These determinations
are consistent with the SM values $S=T=U=0$. The result for S indicates
that negative contributions to this parameter  can remove  the SM 
constraint on $M_H$. For $M_H=600\,GeV$ and $S>0$, as is appropriate in
the simplest technicolor models, $S\leqslant 0.09$, which rules out 
models of this type involving many techni-doublets. These bounds 
can be evaded in models of walking technicolor, where S may be small or
negative. For $T=U=0$, a fit to the S parameter excludes an additional
generation of degenerate chiral fermions at the $99.6\%$ CL. Allowing
$ T = 0.18\pm0.08$, the CL becomes $97\%$. This exclusion of an additional
generation is generally consistent with the number of light neutrinos
$N_{\nu}=2.985\pm0.008$, obtained from the invisible width of the Z boson.
An alternative formulation to S,T, and U, involves the $\epsilon_i$ parameters,
defined in terms of physical observables, $m_W$, $\Gamma_l$, $A^{o,l}_{FB}$,
and $\Gamma_{b\bar{b}}$ \cite{ALT}.

\section{Additional \boldmath{$Z'$} and excited \boldmath{$W^{\pm*}$} Bosons; Trilinear Gauge Couplings}
Additional $Z'$ bosons appear naturally in many Grand Unified Theories
(GUTs) and superstring-inspired models. Frequently discussed examples
are $Z_{\chi}$ of $SO(10)\to SU(5)\times U(1)_{\chi}$, $Z_{\psi}$ of
$E_6 \to SO(10)\times U(1)_{\psi}$, $Z_{\eta}$ of
$E_6 \to SU(5) \times U(1)_{\eta}$, $Z_{LR}$ of $SU(2)_L \times SU(2)_R \times
U(1)$. The $Z_{\eta}$ is a linear combination of $Z_{\chi}$ and  $Z_{\psi}$.
In previous analyses \cite{GAPP}, it was found that the mixing angles
between Z and Z' are severely constrained and that the $95\%$ CL lower
bounds were several hundred GeV, except for $Z_{\psi}$, which contains only
axial vector couplings to ordinary fermions. Furthermore, in certain models in
which the Higgs U(1)' quantum numbers are specified, the lower bounds were 
pushed into the TeV region. Very recently \cite{EL} it has been argued
 that, because of the deviations in $\sigma_{had}$ ($1.7\sigma$) and
$Q_W(Cs)$ ($2.3 \sigma$) relative to the SM predictions, a better fit
to the data can be actually achieved by assuming the existence of an    
additional Z'. In the $Z_{\chi}$ scenario, these authors find 
$M_{Z'}= 812^{+339}_{-152}\,GeV$, $M_H= 145^{+103}_{-61}\,GeV$, a mixing
angle of $O(10^{-3})$ and $\chi^2/d.o.f. = 35/35$. In the $Z_{LR}$ case,
the predictions are $M_{Z'}= 781^{+362}_{-241}\,GeV$, and 
 $M_H= 165^{+155}_{-91}\,GeV$, while in a 
class of $E_6$ models the results are  $M_{Z'}= 287^{+673}_{-101}\,GeV$ and
 $M_H= 101^{+57}_{-39}\,GeV$. The order of magnitude of the mixing angles
and the confidence level in the last two cases are similar to those
found in the $Z_{\chi}$ analysis.\\
Another interesting new physics scenario involves excited $W^{\pm*}$ bosons
that may arise as Kaluza-Klein excitations in certain theories with 
additional compact dimensions, or in models with composite gauge bosons.
Assuming couplings identical to those of the SM $W$, direct searches at
the Tevatron lead to $m_{W*} > 720 \,GeV$ at $95\%$ CL. These excited
$W^{*}$ contribute to muon decay, so that the amplitude for this process 
is proportional to $<m^{2}_{W}>^{-1} = (m^{2}_{W})^{-1} +
 (g^{*}_{2}/g_2)^2\,(m^{*\,2}_{W})^{-1} + \ldots$. Assuming $M_H \lesssim 
200\,GeV$, and using the theoretical expression involving $\Delta\hat{r}$,
one can determine  $<m^{2}_{W}>^{-1}$ from muon decay \cite{MAR}.
 Comparing this result
with the measured value $(m_W)_{exp}$, one finds $m_{W*} > 2.9 (g^{*}_{2}/g_2)
\,TeV$ at $95\%$ CL. This suggests that, in these theories,
the radius of the extra dimension is $R \thickapprox  1/m_{W*} < 10^{-17}
(g_2/g^{*}_2)$~cm. Of course, the bound is significantly relaxed if
 $g^{*}_{2} << g_2$.\\
A very important verification of the SM involves  the measurement of the
trilinear couplings of  $W^{\pm}$ with the photon and Z boson.
 Under a number of theoretical  assumptions, the deviations
from the SM couplings can be parametrized in terms of three quantities,
$\Delta\kappa_{\gamma}$, $\lambda_{\gamma}$, and $\Delta g^{Z}_1$. The
first two contribute to the magnetic and quadrupole moments of 
 $W^{\pm}$ according to the relations: $\mu_W = (e/2m_W)
 (1 + \kappa_{\gamma} + \lambda_{\gamma})$ and $Q_W = -(e/2m_W)
(\kappa_{\gamma} - \lambda_{\gamma})$. The SM predictions are
 $\kappa_{\gamma}=1$, and $\lambda_{\gamma}=0$, so that one defines
$\Delta\kappa_{\gamma} = \kappa_{\gamma} - 1$.            Setting two of the
parameters to zero and fitting the third, recent
measurements at LEP-II \cite {GAPP} give $\Delta\kappa_{\gamma}= 0.038^{+0.079}_{-0.075}$,
$\lambda_{\gamma} = -0.037^{+0.035}_{-0.036}$,
 $\Delta g^{Z}_1 = -0.010\pm0.033$, which are consistent  with the null prediction of the SM.

\section{Electroweak Baryogenesis}
The current consensus is that it is not possible to explain baryogenesis
within the framework of the SM. In the usual scenario, involving the
formation of expanding bubbles with a broken phase, necessary conditions
are sufficiently strong CP violation, a first-order electroweak phase
transition, and the inequality $v(T_c) > T_c$ where $T_c$ is the critical
temperature for the transition and $v(T_c)$ the electroweak scale at that
temperature. These conditions are met only for values of $M_H$ substantially
smaller than the current lower bounds. Thus, in order to explain baryogenesis
in the current scenario, one must invoke physics beyond the SM. In  the MSSM,
 the conditions can be met
if $m_h \lesssim 105\,GeV$ for $m_{\tilde{t}_L}\lesssim 1\,TeV$, or
$m_h \lesssim 115\,GeV$ if $m_{\tilde{t}_L}$ reaches the few TeV region 
\cite{QUIROSSECO}. In both cases the right-handed stop mass is
restricted to lie in the range $100\,GeV\lesssim m_{\tilde{t}_R}\lesssim m_t$.
Thus, in the MSSM there is a ``window of opportunity''.  On the other hand,
it is worth noting that the current lower limit on $M_H$ has recently reached
the $105 \,GeV$ level!

\section{\boldmath{$g_{\mu} - 2$}}
The current SM prediction is  
$a^{SM}_{\mu} = 116591596(67) \times 10^{-11}$, while the experimental
value is $a^{exp}_{\mu} = 116592350(730) \times  10^{-11}$ \cite{CZARNECKIMAR}.
 Thus, the
 difference is  $a^{exp}_{\mu} - a^{SM}_{\mu} = (754\pm733) \times
10^{-11}$ 
At the two-loop level, the electroweak contribution amounts to 
$a^{EW}_{\mu}= 151(4) \times 10^{-11}$, and the hadronic part of the QED
contribution is $a^{had}_{\mu} = 6739(67) \times 10^{-11}$. The present 
goal is to reduce errors to the $40 \times  10^{-11}$ level. $a_{\mu}$ is
sensitive to several types of new physics \cite{CZARNECKIMAR}.
Of particular interest is the SUSY contribution
$a^{SUSY}_{\mu}\thickapprox 140\times 10^{-11} (100\,GeV/\tilde{m})^2\,\tan
\beta$, arising from $\tilde{\nu} \tilde{\chi}^{-}$ and 
 $\tilde{\mu} \tilde{\chi}^{o}$ loops, which will probe the large $\tan\beta$
scenario.

\section{Conclusions}
i) At present levels of accuracy, the SM describes very well the results of
a large number of experiments  ranging from the atomic scale to
about $200\,GeV$.\\
ii) Simple renormalization frameworks have been developed since $\thickapprox
1980$ and  have been applied systematically to study the electroweak 
corrections to a large variety of allowed processes. They play an important
role in current analyses.\\
iii) Two-loop corrections to the muon lifetime in the  V-A Fermi theory
have been completed.\\
iv) The theoretical analysis of the resonance propagators (line shape) sheds
light on the concepts of mass and width of unstable particles. In particular,
in the perturbative regime, the conventional expressions are valid in next
to leading order, but not beyond.\\
v) Theory-driven calculations of $\Delta \alpha ^{(5)}_{h}$ claim to reduce
the error by a factor $\thickapprox 4$. However, the $95\%$ upper bound
$M^{95}_{H}$ is rather insensitive to the change,  because of a partial
cancellation between the reduced error and a shift to larger $M_H$ central
values.\\
vi) The evidence for electroweak corrections beyond the running of $\alpha$
and for electroweak bosonic corrections has become  very sharp:
 $\gtrsim 10\,\sigma$!\\
vii) Trilinear gauge couplings agree very well with SM predictions.\\
viii) With the discovery of the top quark, efforts to constrain $M_H$ have
become increasingly interesting.\\
ix) In particular, the incorporation of the $O(\alpha^2  m_t^2/m_W^2)$ 
corrections has significantly decreased the scheme dependence (theoretical
error) of the
calculations and reduced $M^{95}_{H}$ by $\thickapprox 30\%$!\\ 
x) Simple and accurate formulae for $\sin^2\theta^{lept}_{eff}$ and $m_W$
are available.\\
xi) A major task for the future is the full evaluation of two-loop
 contributions
to $\Delta r, \Delta\hat{r}, \Delta\hat{k}, \ldots $. This would be crucial
if $\delta s^{2}_{eff} \rightarrow 0.01\%$ at the NLC!\\
xii) The present Erler-Langacker global fit to the electroweak data leads to
 $M^{95}_{H}= 235\,GeV$, taking into account the constraint from the direct
searches of H. It has $\chi^2/d.o.f. = 42/37$. Only two observables deviate
from the SM predictions by more than $2\sigma$, with two others at the
$\thickapprox 1.8\sigma$ level. But, the combined $A_b$ differs by
$-2.7\sigma$. It has been argued that this may be due to new physics coupled 
to third generation, in which case new sizeable, tree-level contributions
 to
$g^{b\bar{b}}_{R}$ are required.\\
xiii) Supersymmetry: the precision data is consistent with grand-unification
of
couplings at $\thickapprox 2 \times 10^{16} \,GeV$. A major prediction of the 
MSSM is $m_h \lesssim 135\,GeV$
(and less for small $\tan \beta$). The current consensus is that 
Electroweak Baryogenesis is not feasible in the SM, but there remains a
``window of opportunity'' in the MSSM: $m_h\lesssim 105-115\, GeV$.\\
xiv) There are sharp constraints for non-decoupling new physics:
sequential generations, simplest QCD-like Technicolor models are disfavored
with high probability.\\
xv)  Previous analyses led to lower bounds for masses
of additional $Z'$ bosons and sharp constraints on mixing angles. However, 
it has been recently argued that, because of the current deviations
in $\sigma_{had}$ and $Q_W(Cs)$, a better fit to the precision data can
be actually achieved by postulating the presence of an additional Z'. There
 are also
interesting new bounds on excited $W^{\pm*}$ bosons that may arise in
certain theories with additional dimensions or in composite models.\\
xvi) $g_{\mu} - 2$ is sensitive to SUSY contributions in the large
$\tan \beta$ scenario.\\
xvii) Notwithstanding the current successes of the SM, there remains a plethora
of unsolved, fundamental problems: the precise mechanism of symmetry breaking,
the explanation of the mass spectrum and the number of generations, the
unification with gravity, baryogenesis, the detailed understanding of CP
violation, and probably others that will unfold as we continue our quest.

\section{Acknowledgments}
The author is indebted to J.~Erler, S.~Heinemeyer, P.~Langacker, G.~Quast,
and G.~Weiglein for very useful communications. This work was supported in
part by NSF Grant No. PHY-9722083.


\end{document}